\documentclass[journal]{IEEEtran}
\usepackage{cite}
\usepackage{bm,graphicx}
\usepackage[cmex10]{amsmath}
\usepackage{array}
\usepackage{amssymb,epsfig}

\usepackage{multirow}
\usepackage{tikz}
\usepackage{balance}
\usetikzlibrary{shapes,arrows}

\usepackage{mdwmath}
\usepackage{mdwtab}

\ifCLASSINFOpdf
\else
\fi

\begin{document}

\title{Binary CPMs with Improved Spectral Efficiency}

\author{\IEEEauthorblockN{Malek Messai,~\IEEEmembership{Student Member,~IEEE}, Amina Piemontese,~\IEEEmembership{Member,~IEEE}, 
Giulio Colavolpe,~\IEEEmembership{Senior Member,~IEEE}, Karine Amis,~\IEEEmembership{Member,~IEEE}, 
and Fr\'ed\'eric Guilloud,~\IEEEmembership{Member,~IEEE} }

  \thanks{Manuscript received August 18, 2015; revised September 25, 2015 and accepted November 12, 2015. The associate editor coordinating the review of this letter and approving it for publication was Zhaocheng Wang.}

\IEEEcompsocitemizethanks{M. Messai, K. Amis and F. Guilloud are with the Signal and Communication Department of Telecom Bretagne-Institut 
Telecom, CNRS Lab-STICC (UNR 6285), Brest 29200, France (e-mail:{malek.messai, karine.amis, frederic.guilloud}@telecom-bretagne.eu). 
A. Piemontese is with the Department of Signals and Systems, Chalmers University of Technology, Gothenburg, 
Sweden (e-mail: aminap@chalmers.se). G. Colavolpe is with Universit\`a di Parma, Dipartimento di Ingegneria dell'Informazione, Viale G. P. 
Usberti, 181A, I-43124 Parma, Italy (e-mail: giulio@unipr.it).}

   }

\maketitle

\begin{abstract}
We design new continuous phase modulation (CPM) formats which are based on the combination of a proper precoder with binary input and a ternary CPM. The proposed precoder constrains the signal phase evolution in order to increase the minimum Euclidean distance, and to limit the bandwidth expansion due to the use of a ternary CPM. The resulting schemes are highly spectrally efficient and outperform classical binary and quaternary formats in terms of coded and uncoded performance.
\end{abstract}
\vspace*{-0.2cm}

\IEEEpeerreviewmaketitle

\section{Introduction}
Continuous phase modulations (CPMs) are advanced modulation formats which are often employed in satellite communications thanks to their immunity against nonlinear distortions, stemming from the constant envelope, their claimed power and spectral efficiency, and their recursive nature which allows to employ them in serially concatenated schemes. CPMs have been included in the DVB-RCS2 standard~\cite{DVB-RCS2}, with the aim of enabling the use of cheaper amplifier components in modems, and hence lower cost terminals.

The spectral efficiency (SE) is one of the main figures of merit for satellite communications. In this letter, we design new binary schemes with a
much larger SE than classical CPMs. They are based on the combination of a proper precoder with binary input and a ternary CPM format.

The use of a precoder in conjunction with a CPM modulator is rarely considered in the literature. To our knowledge, it has been recently investigated to improve the robustness of classical binary CPMs to a modulation index mismatch in~\cite{MeCoAmGu15b} and to have a simple (albeit suboptimal) symbol-by-symbol detection architecture in shaped-offset quadrature phase-shift keying (SOQPSK) modulation~\cite{PeRi07}. 
In~\cite{Fo91}, an improved \mbox{\textit{multi-h}} CPM is proposed which increases the minimum Euclidean distance of binary CPMs but without taking care of the bandwidth. The state transitions of the phase trellis are controlled by employing a set of modulation indexes in the same symbol interval, depending on the information symbol to be transmitted.

This paper is organized as follows. In Section~\ref{sec:cpm_sig} we introduce the system model. The proposed schemes and the corresponding
detectors are described in Section~\ref{sec:pro_det}. Numerical results are reported in Section~\ref{num_res}, where we show how the proposed schemes outperform binary and quaternary CPMs both in terms of uncoded performance and SE. Finally, conclusions are drawn in Section~\ref{conc}.

\section{System Model}
\label{sec:cpm_sig}
The complex envelope of a CPM signal can be written as
\begin{equation}\nonumber
s(t,{\bm a})=\sqrt{\frac{E_s}{T}} \exp \Big\{\jmath 2 \pi h \sum_n a_n q(t-nT) \Big\},   \label{e:bdb_sig}
\end{equation}
where $E_s$ is the average energy per information symbol, $T$ the symbol interval, $h=r/p$ the modulation index ($r$ and $p$ are relatively prime integers), $q(t)$ is the {\em phase-smoothing response}, and its derivative is the {\em frequency pulse}, assumed of duration $LT$ and area 1/2. The symbols ${\bm a}=\{a_n\}$ take on values in the alphabet $\{ \pm 1, \pm 3 \cdots \pm (M-1)\} $ if $M$ is even, and in the alphabet $\{ 0, \pm 2, , \pm 4\cdots \pm (M-1)\} $ if $M$ is odd.

In the  generic time interval~$[nT,nT+T)$, the CPM signal is completely defined by symbol~$a_n$ and state $\sigma_n=(\omega_n,\theta_n)$~\cite{Ri88}, where 
\begin{equation}\nonumber
  \omega_n = (a_{n-1},a_{n-2},\dots,a_{n-L+1})
\end{equation}
is the correlative state and $\theta_n$ is the phase state which takes on $p$ values evenly spaced around the unit circle, and can be recursively defined as
  \begin{equation}\nonumber
  \theta_n = [\theta_{n-1} + \pi h a_{n-L} ]_{2\pi}\, ,
  \label{e:phase_state}
 \end{equation}
where $[\cdot]_{2\pi}$ denotes the ``modulo $2\pi$'' operator. 

\begin{figure}[!t]
 \center
 \includegraphics[width=5cm]{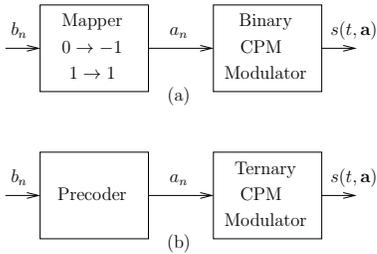}
  \vspace{-2mm}
 \caption{Compared schemes. (a) Classical binary CPM. (b) Proposed scheme.}
 \label{fig:schemes}
 \vspace{-5mm}
 \end{figure}

Our aim is to improve the SE of classical binary CPM schemes through the design of a suitable precoder. A classical binary CPM modulator, can be represented as depicted in Fig.~\ref{fig:schemes}(a). Information bits $\{b_n\}$, belonging to the alphabet $\{0,1\}$, are directly mapped into symbols $\{a_n\}$ belonging to the alphabet $\{\pm 1\}$, and then go at the input of a binary CPM modulator.
The scheme proposed in this paper is instead  based on the combination of a precoder, which receives at its input bits $\{b_n\}$ belonging to the alphabet $\{0,1\}$ and provides at its output ternary symbols $\{a_n\}$ belonging to the alphabet $\{0,\pm 2\}$, and a ternary CPM. This precoded scheme is shown in  Fig. \ref{fig:schemes}(b).

We consider a transmission over an additive white Gaussian noise (AWGN) channel. The complex envelope of the received signal thus reads
\begin{equation}
r(t) =   s(t,{\bm a}) + w(t)   \, ,\label{e:r(t)}
\end{equation}
where $w(t)$ is a complex-valued white Gaussian noise process with independent components, each with two-sided power spectral density $N_0/2$.

\section{Proposed Precoder and Corresponding Detector}
\label{sec:pro_det}
The SE is defined as 
\begin{equation}
\text{SE}=\frac{I}{BT}\,\quad[\textrm{bit}/\textrm{s}/\textrm{Hz}] \, ,  \label{eq:se}
\end{equation}
where $I$ is the information rate (IR), i.e., the amount of information transmitted per channel use, and $B$ is the bandwidth occupied by the transmitted signal. This normalization is required to capture the different bandwidth occupancy of different modulation formats. In other words,  by considering the SE, i.e.,  the amount of information transmitted per unity of time and per unity of bandwidth,  we are also considering the effect of the precoding on the bandwidth occupancy. 

A simple way to increase the IR is to increase the alphabet size, but this results in an expansion of the spectrum occupancy and does not guarantees an overall improvement of the SE. Moreover, this leads to an increase in the computational complexity of the detection algorithm.

The design of the proposed precoder is based on the increase of the minimum Euclidean distance, which dominates the performance at high signal-to-noise ratio (SNR) and is hence closely related to the IR. Furthermore, the proposed scheme is also able to limit the spectral occupancy increase due to the use of a ternary modulation instead of a binary one. This is based on the observation that the width of the main spectral lobe is related to the phase hopping that becomes important with the increase of $M$ or $h$ \cite{AnAuSu86}. 
In the case of ternary CPMs, this spectral widening is mainly due to the transitions from 2 to -2 and vice versa. 
In summary, the challenge is to design a precoder in such a way to ensure simultaneously: 
\begin{itemize}
 \item an increase of the minimum distance (in order to increase the IR);
 \item the absence of transitions from 2 to -2 and vice versa (in order to limit the bandwidth expansion).
\end{itemize}
To design a precoder which increases the minimum Euclidean distance, we consider the upper bound given by the distance between paths in the phase tree which have a common starting phase and have a merge after a finite number of time intervals~\cite{AnAuSu86}. The idea is to design a precoder which can delay the first inevitable merge. This is the same principle used for the \textit{multi-h} CPMs~\cite{AuSu82,Fo91}.

\begin{figure}[!t]
\center
\includegraphics[width=7.8cm]{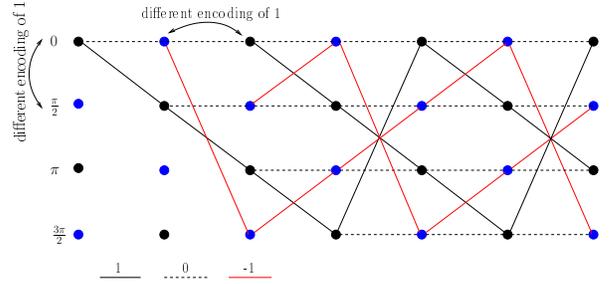}
\caption{The phase tree modulo 2$\pi$ of a full response CPM modulation with $h$=1/4  and REC frequency pulse (the proposed scheme).}
\label{fig:trellis_h14}
\vspace{-5mm}
\end{figure}

% The phase state \textcolor{red}{$\theta_n$} can assume the following $p$ values, evenly spaced around the unit circle
% $$
% \Big\{0,2\pi h,\dots, \big(p - 1\big)2 \pi h \Big\}\,.
% $$
The joint phase tree of the precoder/modulator is time varying and we denote by $\varphi_n$ the phase state in the $n$th symbol interval and by $\varphi_n(i)$ the $i$th value that it can assume.\footnote{Without loss of generality, we assume that the states values are ordered in an increasing order.} 
We introduce the proposed precoded CPM with an example. We consider the phase tree for a CPM with $h = \frac{1}{4}$ and rectangular (REC) frequency pulse with $L=1$ (full response), which includes the phase state values $\{0, \pi/2, \pi, 3 \pi/2  \}$, and is shown in Fig.~\ref{fig:trellis_h14}. From each state, we have two branches corresponding to the bits ``0'' and ``1''. The bit ``0'' is encoded as 0, whereas the bit ``1'' is encoded as 2 or -2 based on the following rules:
\begin{itemize}
 \item two adjacent states $\varphi_n(i)$ and $\varphi_n(i+1)$ should have different encoding of the bit ``1'' (\textit{in order to delay the first merge});
 \item also the two successive states $\varphi_n(i)$ and $\varphi_{n+1}(i)$  should have different encoding of the bit ``1'' (\textit{to avoid the alternation $2 \rightarrow-2$ and vice versa, which imply phase transitions causing a bandwidth expansion}). 
\end{itemize}
With this construction, at each time instant $t=nT>(L+1)T$ all phase values in the tree are possible, but for every phase state the encoding of the bit ``1'' in the even-numbered symbol intervals is different from that in the odd-numbered symbol intervals. In this sense the tree is time-varying.

In order to derive the precoder equation, we now consider a possible path in the trellis: at the time instant $t=nT$ the bit ``0'' is encoded as ``0'', while the encoding of the bit ``1'' depends on the encoding of the previous ``1'' and on the number of time intervals $d$ that are between them.
The following precoder satisfies the required constraints for all values of $p>2$ and $L$:
\begin{equation}\nonumber
a_n = b_n a_{n-d} (-1)^{d+1},   \label{e:precod}
\end{equation}
where $b_n$ is the bit transmitted in the symbol interval $n$ and $a_{n-d}$ is the latest symbol such that $|a_{n-d}|=2$.

In the considered example, the minimum distance is governed by the error event ${\bm e}_{min} = (2,-2)$ of length $2T$ for binary CPMs, and by the error event ${\bm \tilde{e}}_{min} = (2,0,-2)$ of length $3T$ for the proposed precoded scheme. The multiplicity of the error events with minimum distance is the same in both cases. In general, the first merge in the phase tree occurs at time $t=(L+1)T$ for the binary CPM case and at time $t=(L+2)T$ for the proposed scheme when $p>2$.  
% \subsection{Corresponding Detector}
% \label{sec:detector}

We now consider the corresponding detection algorithms, which are described by means of the detector phase trellis. In each time interval, although ternary, the transmitted symbol is drawn from one of two binary alphabets $\{-2, 0\}$ or $\{0, 2\}$. 
When $h$ has an even denominator, the detector trellis can be directly derived from the phase tree, and detection can be performed with the same complexity of the corresponding binary CPM.
% For example, the reception trellis for $L = 1$ and $h = 1/4$ is illustrated in Fig.~\ref{fig:trellis_h14}.

\begin{figure}[!t]
\center
\includegraphics[width=7.5cm]{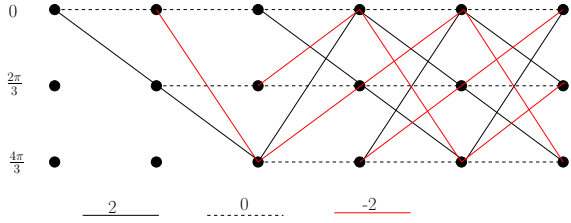}
\caption{The phase evolution of a full response CPM modulation with $h$=1/3  and REC frequency pulse (odd modulation index denominator).}
\label{fig:tree_h13}
\vspace{-3mm}
\end{figure}

\begin{figure}[!t]
\center
\includegraphics[width=7.8cm]{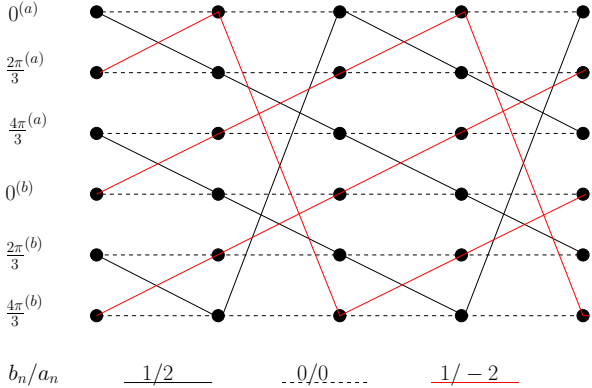}
\caption{The proposed phase trellis for odd modulation index denominator.}
\label{fig:trellis_h13}
\vspace{-5mm}
\end{figure}

When $p$ is odd, the phase evolution at the transmitter can no more be used to derive the detector trellis since we do not have a one-to-one correspondence between the transmitted bit sequences and phase trajectories. The phase evolution for the case of a full response CPM with REC frequency pulse and $h = 1/3$ is shown in Fig.~\ref{fig:tree_h13}. Even if the transmitted symbol is drawn from one of two binary alphabets $\{-2, 0\}$ or $\{0, 2\}$, each state in the phase tree has three outgoing edges: we can not associate a transmitted bit sequence to each phase trajectory. This can be overcome by properly constructing the trellis with an even number of phase states equal to $2p$. For each physical phase state value $\phi\in \{ 0,2\pi h,\dots, \big(p - 1\big)2 \pi h\}$, we introduce $\phi^{(a)}$ and $\phi^{(b)}$ which have different encoding of the bit ``1''.
% When $p$ is odd, the constraint on the different encoding of the bit ``1'' between two adjacent states leads to a reception trellis with an even number of phase states equal to $2p$. 
Then, for the proposed scheme $\varphi_n$ belongs to $\{0^{(a)}, {2\pi h}^{(a)},..,{2 \pi (p-1)h}^{(a)},0^{(b)}, {2\pi h}^{(b)},.., {2 \pi (p-1)h}^{(b)}\}$. In Fig.~\ref{fig:trellis_h13}, we show the detector trellis for the scheme of Fig.~\ref{fig:tree_h13}. For this example, in the first symbol interval the edge going out from the state $0^{(a)}$ encodes the bit ``1'' as 2, while that from the state $0^{(b)}$ encodes it as $-2$. The role of the two states changes in the following symbol interval. It is worth noticing that the derived trellis satisfies the required constrains on the encoding of the bit ``1'' between adjacent and successive states.
% The corresponding state diagram of the overall precoded scheme is shown in Fig.~\ref{fig:diag_pha_state}.
% \begin{figure}[!t]
% \center
% \includegraphics[width=8.5cm,height=5.5cm]{figures/diag_pha_state_A}
% \caption{Phase state diagram in the case of $h=1/3$ (odd modulation index denominator).}
% \label{fig:diag_pha_state}
% \end{figure}

% It is important to mention that when $p$ is even, $\theta_i^{+}(n)$ and $\theta_i^{-}(n)$ will have the same encoding of 1 and then the branch metrics going out from these two states will be equal. This implies that the two state $\theta_i^{+}(n)$ and $\theta_i^{-}(n)$ can be  combined into one state $\theta_i(n)$. 
% The number of CPM states $\sigma_n$ of the proposed scheme, which is related to the computational complexity of the receiver, is $p (\left[p \right]_{2}+1) 2^{L-1}$, whereas for classical CPM it is $p M^{L-1}$.
% Thereafter, there will be two different trellises: the first one is exactly the same as in the binary CPM case when $p$ is even and the second will have double \emph{phase state} number when $p$ is odd.
\section{Simulation Results} \label{num_res}
In this section, we consider the performance of the proposed scheme, compared to classical binary and quaternary CPMs. We first evaluate the bit error rate (BER) in uncoded transmissions. Then, we consider the performance in terms of SE, which in practice can be achieved by adopting a suitable channel code. 
Due to lack of space, we limit our results to CPM formats with REC frequency pulse. In particular, we consider frequency pulse of length $T$ (1REC) and  $2T$ (2REC).
\subsection{Uncoded Performance}
We consider the uncoded performance of the proposed system in case of transmission over an AWGN channel. Particularly, we evaluate the asymptotic performance for high values of $E_b/N_0$, where $E_b$ is the mean energy per information bit. Let us denote by ${\bm e}={\bm a}-\hat{\bm a}$ the sequence representing the difference between the transmitted sequence ${\bm a}$ and the erroneous one $\hat{\bm a}$. Without loss of generality, we will assume that any considered error event starts at time $n=0$.
We will also denote the normalized squared Euclidean distance as~\cite{AnAuSu86}
\begin{equation}\nonumber
d^2({\bm e}) = \frac{1}{E_b} ||s(t,{\bm a}) - s(t,\hat{{\bm a}}) ||^2\,. \label{eq:nor_dis}
\end{equation}

The BER for the optimal maximum a posteriori probability (MAP) sequence detector is well approximated by~\cite{PeRi05}
\begin{equation}\label{eq:prob_error}
P_b \approx \frac{  n_{ {\bm e}_{min} }   m_{ {\bm e}_{min} }    }{\log_2( S)  S^{ R_{ {\bm e}_{min} }  }}  {\rm Q}\left( \sqrt{d_{min}^2 E_b/N_0}\right)\, ,
\end{equation} 
where $S$ is equal to $M$ for classical CPMs and is equal to 2 for the proposed scheme, $d_{min}=\min_{\bm e} d({\bm e})\,, {\bm e}_{min}=\arg\!\min_{\bm e} d({\bm e})\,,$
$n_{ {\bm e}_{min}}$ is the number of bit errors (i.e., on the sequence $\{b_n\}$) caused by the error event ${\bm e}_{min}$, $m_{{\bm e}_{min}} = 2 \prod_{i=0}^{R_{{\bm e}_{min}}-1} (S-\frac{|e_{min,i}|}{2})$,  $R_{{\bm e}_{min}}$ is the span of symbol times where ${\bm e}_{min}$ is different from zero and ${\rm Q}(x)$ is the Gaussian Q function.
% If there are more sequences ${\bm e}$ corresponding to  $d_{min}$, the bit error probability will have more terms of the form of the right hand side of (\ref{eq:prob_error}), each one corresponding to a different sequence ${\bm e}$. 
% The coefficient $\frac{  n_{ {\bm e}_{min} }   m_{ {\bm e}_{min} }    }{\log_2 (S) S^{ R_{ {\bm e}_{min} }  }}$ is often called \textit{multiplicity} of the error event with minimum distance.
Now, it only remains to identify the error events corresponding to $d_{min}$. This can be done by working on the phase tree, as described in~\cite{AnAuSu86}. 
% We consider different modulation formats and compute the corresponding parameters $n_{ {\bm e}_{min} }$, $m_{ {\bm e}_{min} }$,  $R_{ {\bm e}_{min} }$ and $d_{min}$. 

For fixed $h$ indexes and frequency pulses, we compute~(\ref{eq:prob_error}) for several classical binary and quaternary schemes and for the proposed one. The results are summarized in Table~\ref{tab1}, where we report $d^2_{min}$ and the gain in dB that the proposed scheme achieves at BER=$10^{-4}$ with respect to the classical formats. The BER simulations, not reported here for a lack of space, confirm the analytic results, and show the large performance gain of the proposed scheme with respect to the classical CPMs for the cases listed in Table~\ref{tab1}.

\begin{table}[!t]
\centering
\caption{Uncoded performance.}
 \vspace*{-0.2cm}
\label{tab1}
\begin{tabular}{|l|l|l|l||l|l|l|l|}
\hline
 \multicolumn{2}{|c|}{ Scheme} & $d^2_{min}$ & Gap &\multicolumn{2}{c|}{ Scheme} & $d^2_{min}$ & Gap \\ \hline
\multirow{2}{*}{\begin{tabular}[c]{@{}l@{}}1REC\\ $h$=1/4\end{tabular}} & bin.  & 0.726  & 3.85dB & \multirow{2}{*}{\begin{tabular}[c]{@{}l@{}}2REC\\ $h$=1/4\end{tabular}}  & bin.  & 0.492 & 4.7dB \\ 
                                                                        & quat. & 1.453 & 0.81dB & &quat. &0.984 &1.75dB\\ 
                                                                        & prop.   & 1.726  & - & &prop. &1.453&-\\ \hline
\multirow{2}{*}{\begin{tabular}[c]{@{}l@{}}1REC\\ $h$=1/5\end{tabular}} & bin.  & 0.486  & 3.85dB &\multirow{2}{*}{\begin{tabular}[c]{@{}l@{}}2REC\\ $h$=1/5\end{tabular}}  & bin.  & 0.32 &4.85dB\\ 
                                                                        & quat. & 0.972  & 0.9dB & & quat.&0.64&1.9dB\\ 
                                                                        & prop.   & 1.177  & - &  &prop.& 0.972&-\\ \hline
\multirow{2}{*}{\begin{tabular}[c]{@{}l@{}}1REC\\ $h$=2/7\end{tabular}} & bin.  & 0.913  & 3.7dB & \multirow{2}{*}{\begin{tabular}[c]{@{}l@{}}2REC\\ $h$=2/7\end{tabular}}  & bin.  & 0.634 &4.6dB\\ 
                                                                        & quat. & 1.827  & 0.75dB & &quat.&1.268&1.65dB\\ 
                                                                        & prop.   & 2.136  & - &  &prop.&1.827&-\\ \hline
\end{tabular}
 \vspace{-5mm}
\end{table}

\subsection{Spectral Efficiency}
The CPM bandwidth is theoretically infinite because the power spectral density of a CPM signal has rigorously an infinite support. To define the signal bandwidth, we consider a frequency division multiplexed (FDM) system where each user employs a CPM. In order to avoid boundary effects, we assume a system with an infinite number of users which transmit at the same power, employ the same modulation format and are equally spaced in frequency. Under these conditions, the frequency spacing is a measure of the signal bandwidth. Particularly, we let the user interfere and define the system bandwidth as the frequency separation which maximizes the SE achievable by a single user detector. The reader is referred to~\cite{BaFeCo09} for further details on this bandwidth definition.

To compute $I$ in~(\ref{eq:se}), we can use the simulation-based technique described in \cite{ArLoVoKaZe06}, which only requires the existence of an optimal MAP symbol detector for the channel~(\ref{e:r(t)}). This problem is an instance of mismatched detection~\cite{MeKaLaSh94}, since we are interested in the computation of the IR when the actual channel is a FDM system with interfering users, while the receiver is a single user receiver which is designed according to the channel~(\ref{e:r(t)}).
The simulation-based method described in \cite{ArLoVoKaZe06} allows to evaluate the achievable IR as 
\begin{equation}
I=E\left\{ \!\log\frac{p(\boldsymbol{r}|\boldsymbol{a})}{p(\boldsymbol{r})}\!\right\}\,, \quad[\textrm{bit}/\textrm{ch.use}]\, , \label{eq:AIR}
\end{equation}
where $\boldsymbol{r}$ is a suitable vector of sufficient statistics extracted from the continuous-time received signal.
In~(\ref{eq:AIR}), the probability density functions $p(\boldsymbol{r}|\boldsymbol{a})$ and $p(\boldsymbol{r})$ can be evaluated recursively through the forward recursion of the optimal MAP symbol detection algorithm for the channel~(\ref{e:r(t)}). 
% This receiver can assure communication with arbitrarily small error probability when the transmission rate at the CPM modulator input does not exceed $I$ bits per channel use, provided that a suitable channel code is adopted.

\begin{figure}[!t]
\center
\includegraphics[width=8.8cm]{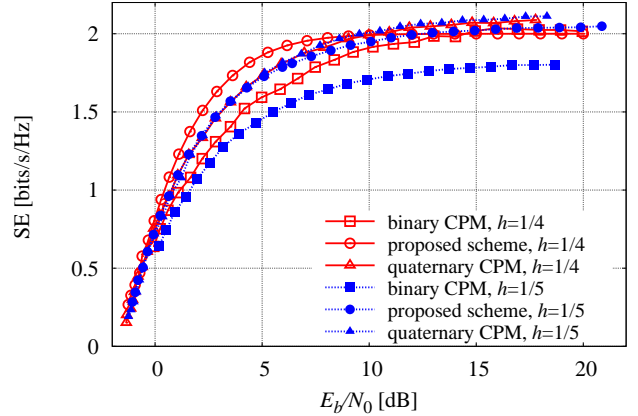}
 \vspace{-10mm}
\caption{Spectral efficiency for 1REC modulation formats.}
\label{fig:SE_1REC}
 \vspace{-5mm}
\end{figure}

\begin{figure}[!t]
 \center
 \includegraphics[width=8.8cm]{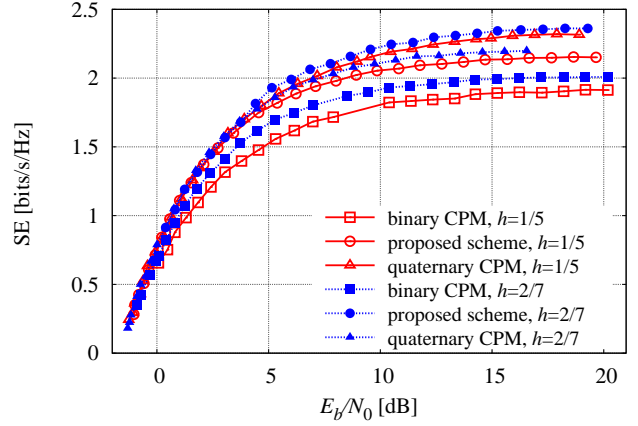}
  \vspace{-10mm}
 \caption{Spectral efficiency for 2REC modulation formats.}
 \label{fig:SE_2REC}
  \vspace{-5mm}
 \end{figure}

We plot the optimized SE \cite{BaFeCo09} as a function of $E_b/N_0$. The performance of the proposed scheme is compared to binary and quaternary CPMs. 
Fig.~\ref{fig:SE_1REC} shows the SE of a full response CPM format (1REC). It can be observed that the precoded scheme results in higher SEs than the binary CPMs, and compared to the quaternary formats has better performance for $h=1/4$ and the same performance $h=1/5$. 
The same conclusion is drawn in the case of the 2REC formats considered in Fig.~\ref{fig:SE_2REC}.  
We observe that the proposed scheme performs better than the binary case. Compared to the quaternary CPM formats, it performs better in the case $h=2/7$ and allows to reach the same performance for $h=1/5$. 
In some cases, the quaternary formats allow to achieve higher asymptotic SE, but the corresponding SNR region has no practical interest.

Finally, we compare the considered schemes in terms of the computational complexity of the corresponding detectors. In the case of classical CPMs, the optimal detector has a computational complexity which is proportional to $pM^L$, while the optimal detector for the proposed scheme has a complexity proportional to  $p 2^L$ if $p$ is even, and to $p 2^{L+1}$ if $p$ is odd. Therefore, compared to the binary CPMs, the proposed scheme has the same receiver complexity if $p$ is even, while the complexity increases in the case of odd values of $p$. Interestingly, the proposed scheme has always less or equal complexity than the quaternary formats since:
\begin{equation}\nonumber
\underbrace{p \times (\left[p \right]_{2} +1) \times 2^L}_{\mbox{Prop. sch. compl.}} \leq  \underbrace{p \times 4^L}_{\mbox{Quat. CPM compl.}}  , \quad \forall p, \quad \forall  L \label{eq:complexity}
\end{equation}

% 
% \begin{figure}[!t]
% \center
% \includegraphics[width=9.3cm]{figures/BER_1REC}
% \vspace{-7mm}
% \caption{BER performance comparisons for 1REC modulation (simulations  and closed-form asymptotic expressions).}
% \label{fig:ber_1rec}
% \end{figure}

% \begin{figure}[!t]
% \center
% \includegraphics[width=9.3cm]{figures/BER_2REC}
% \vspace{-7mm}
% \caption{BER performance for 2REC modulation formats (simulations and closed-form asymptotic expressions).}
% \label{fig:ber_2rec}
% \end{figure}

\vspace{-2mm}

\section{Conclusions}
\label{conc}
We proposed a new family of binary formats given by the combination of a precoder with a ternary CPM.
The new schemes are very attractive form the SE point of view, achieving better results than classical formats. Moreover, the corresponding detection algorithms have less or equal computational complexity than the quaternary formats.

\vspace{-2mm}


\end{document}